\documentclass[fleqn,usenatbib]{mnras}

\usepackage{newtxtext,newtxmath}

\usepackage[T1]{fontenc}
\usepackage{ae,aecompl}

\usepackage{graphicx}	
\usepackage{amsmath}	
\usepackage{amssymb}

\usepackage{color}
\definecolor{titlecol4}{rgb}{0.039,0.361,0.569}

\title[Galactic Conformity in Morphology]{Galactic Conformity in both Star-formation and Morphological Properties}

\author[Otter et al.]{
Justin A. Otter,$^{1}$\thanks{E-mail: jaotter123@gmail.com}
Karen L. Masters,$^{1}$\thanks{E-mail: klmasters@haverford.edu}
Brooke Simmons$^{2,3}$ and
\newauthor
Chris J. Lintott$^4$
\\
$^{1}$Haverford College, Department of Physics and Astronomy, 370 Lancaster Avenue, Haverford, Pennsylvania 19041, USA\\
$^{2}$Physics Department, Lancaster University, Lancaster, LA1 4YB, UK \\
$^{3}$Center for Astrophysics and Space Sciences (CASS), Department of Physics, University of California, San Diego, CA 92093, US \\
$^4$Oxford Astrophysics, Department of Physics, University of Oxford, Denys Wilkinson Building, Keble Road, Oxford, OX1 3RH, UK
}

\date{Accepted XXX. Received YYY; in original form ZZZ}

\pubyear{2019}

\begin{document}
\label{firstpage}
\pagerange{\pageref{firstpage}--\pageref{lastpage}}
\maketitle

\begin{abstract}
 We investigate one-halo galactic conformity (the tendency for satellite galaxies to mirror the properties of their central) in both star-formation and morphology using a sample of 8230 galaxies in 1266 groups with photometry and spectroscopy from the Sloan Digital Sky Survey, morphologies from Galaxy Zoo and group memberships as determined by Yang et al. This is the first paper to investigate galactic conformity in both star-formation and visual morphology properties separately. We find that the signal of galactic conformity is present at low significance in both star-formation and visual morphological properties, however it is stronger in star-formation properties. Over the entire halo mass range we find that groups with star-forming (spiral) centrals have, on average, a fraction 0.18$\pm$0.08 (0.08$\pm$0.06) more star-forming (spiral) satellites than groups with passive (early-type) centrals at a similar halo mass. We also consider conformity in groups with four types of central: passive early-types, star-forming spirals, passive spirals and star-forming early-types (which are very rarely centrals), finding that the signal of morphological conformity is strongest around passive centrals regardless of morphology; although blue spiral centrals are also more likely than average to have blue spiral satellites. We interpret these observations of the relative size of the conformity signal as supporting a scenario where star-formation properties are relatively easily changed, while morphology changes less often/more slowly for galaxies in the group environment. 
\end{abstract}


\section{Introduction}

 ``Galactic conformity" is a term that was coined to describe the tendency for satellite galaxies in a group or cluster to have similar properties to the central galaxy in their group \citep{weinmann_properties_2006}. 
 
 It has been known for a long time that a range of galaxy properties correlated with environment \citep[e.g.][]{oemler1974,dressler1980,haynes1984}, with younger, more gas rich, bluer, and spiral galaxies tending to avoid dense clusters and groups. It was noted as early as \citet{Hubble1931} that the morphological distribution of galaxies varies with environment; and there have for a long time been observations that show galaxies in close pairs are more likely to have similar colours and morphology than random pairs \citep{holmberg1958,madore1986}; this is one of the ``Holmberg Effects", put down to the impact of interaction in some, but not all pairs \citep{Kennicutt1987}. However the observation of ``galactic conformity" (as first presented in \citealt{weinmann_properties_2006}) suggested that environment, or even halo mass alone wasn't the only factor -- revealing that even at the same halo mass, galaxies in groups with different central galaxy properties would tend to match the central galaxy.  
 
 The original description of galactic conformity focused on the star-formation properties of galaxies. \citet{weinmann_properties_2006} classified their galaxies as ``late" or ``early" on the basis of a combination of colour and specific star-formation rate (sSFR). However, as is quite commonly done, due to their strong correlation \citep[e.g][]{strateva2001}, colour and morphology were assumed to be equivalent, and so the interpretation mixed discussion of effects which impact the star-formation properties (e.g. removal of gas), and those that impact both morphology and star-formation (e.g. merging). 
 
 This has continued to be a tendency in many papers following up, and investigating galactic conformity. For example, \cite{hartley_galactic_2015} studies galactic conformity out to $z\sim 2$. In considering physical origins of the star-formation conformity signal, they also consider processes which impact the gas of a group, such as gas stripping, in tandem with those which impact the morphology of galaxies, like mergers. \cite{prescott_galaxy_2011} investigate colour conformity and similarly conflate definitions of early- and late-type galaxies based on star-formation, with those based on morphology.
 
 Other work has focused primarily on the gas content of galaxies. \cite{knobel_quenching_2015} detects galactic conformity in both the fraction of star-forming satellites as well as the ``quenching efficiency" of the satellites around star-forming and quenched centrals. They match five environmental parameters: the stellar mass of the satellites, halo mass, central stellar mass, local over-density, and group-centric distance and still find this result, and thus conclude that conformity in star-formation properties must be driven by another physical parameter shared by centrals and satellites. \cite{wang_satellite_2012} look at colour conformity in isolated groups, and compare their results with a simulation. They find that the simulation reproduces the star-formation conformity effect, and conclude that the conformity signal is primarily a result of higher halo mass in groups with red centrals, and from differing gas properties at fixed halo mass. \cite{kauffmann_re-examination_2013} use a semi-analytic model to study star-formation conformity, and find they do not reproduce the observational signal of conformity. They go on to propose that this suggests accretion in high mass halos and gas ``pre-heating" in low mass halos are likely drivers of star-formation conformity. However observations that conformity was weaker at higher red-shifts \citep{Berti2017} suggest that a model where large scale tidal fields generate conformity, is more likely. 
 
 There has been only one previous work which investigated galactic morphological conformity explicitly (however even it still mixed in colour for the classification of satellites): \citet{ann_galactic_2008} studies the radial dependence of satellite morphology (using an automated proxy for morphology which depended on galaxy colour, colour gradient and concentration) around central galaxies (with visual morphology classifications). They detect strong ``morphological" conformity (meaning that redder, more concentrated satellites were more likely to be found in groups with visually classified elliptical central galaxies). 
 
 Previous work by \citet{calderon_small-_2018} uses the S\'{e}rsic index as a proxy for morphology. While this proxy is very reasonable statistically speaking, there are high uncertainties in individual cases which could smear out signals due to cross contamination. \citet{Simmons_2018} investigated the reliability of the S\'{e}rsic index as a proxy of morphology in detail. Though the focus of this work was on active galactic nuclei (AGN) host galaxies, the results were shown to hold for a control sample of inactive galaxies as well. They found that  $n < 1.5$ reliably selected disk-dominated galaxies, and that $n > 3$ might reliably select strong bulge galaxies but not necessarily only elliptical galaxies as there could still be significant disks present. They were also unable to select only bulge- or disk-dominated galaxies for intermediate S\'{e}rsic values. Lastly, the S\'{e}rsic index has completeness issues because azimuthally smooth fits are more likely to crash the more ``featured'' (non-axisymmetric) a galaxy is.
 
 In this paper we present the first investigation of one-halo galactic conformity in both star-formation and morphology properties separately, using a sample which has visual morphologies (independent of colour properties) for {\bf all} galaxies (satellites and centrals).
 
 Galaxy morphology and star-formation properties are strongly correlated \citep[e.g.][]{RobertsHaynes1994,Kennicutt1998,strateva2001,Schawinski2014}, however it has been clearly shown that there exist significant populations of red spirals \citep{Bamford2009,Masters2010} and blue ellipticals \citep{Bamford2009,Schawinski2009}, and what's more that the environmental dependence of colour (star-formation properties) and morphology are subtly different \citep{Bamford2009,Skibba2009}, with star-formation properties changing more readily with environment than morphology (e.g. \citep{Skibba2009} show that at fixed colour there is very little trend of morphology with density, while at fixed morphology colour trends with density are strong). This suggests that galactic conformity signals should differ when star-formation and morphology of galaxies are considered separately, a prediction that as yet has not been tested observationally.
 
 In this paper we investigate one-halo galactic morphological and star-formation conformity completely separately,  for the first time in a sample where both the centrals and satellites have visual morphological classifications. We use a similar sample and star-formation measures to \citet{weinmann_properties_2006} and demonstrate that we can reproduce their result. Followup works have cast doubt on whether the conformity signal detected by \cite{weinmann_properties_2006} and others is real, or a systematic introduced by the group finding algorithm \cite[e.g.][]{campbell_assessing_2015,calderon_small-_2018,berti_primus_2019}. In this work we add visual morphologies for all satellite and central galaxies from the Galaxy Zoo project \citep{Lintott2008,Willett2013}, and focus on the relative size of conformity signals.
 
 In Section \ref{sample} we describe our sample selection and source of all data and data products we use. Section \ref{results} describes our results, as well as our work to test how robust they are to group catalogue completeness. We summarize our results and provide conclusions in Section \ref{conclusions}. Where distances are needed we use a Hubble constant of $H_0 = 70$ km/s/Mpc.

\section{Sample Selection and Data}
\label{sample}

The sample of galaxies in this paper is based on galaxies which were part of Galaxy Zoo 2 \citep{Willett2013}, which was selected from the Sloan Digital Sky Survey (SDSS) Data Release 7 \citep{abazajian_seventh_2009}. We limit the volume of our sample with a spectroscopic redshift limit of ${0.01 < z < 0.06}$, and an accompanying absolute magnitude limit of $M_r < -19.37$. The lower redshift limit reduces the impact of peculiar velocities in redshift measurements, and the upper limit is needed to ensure galaxies are close enough to resolve morphological properties in SDSS imaging.

\begin{figure}
\includegraphics[width=0.4\textwidth]{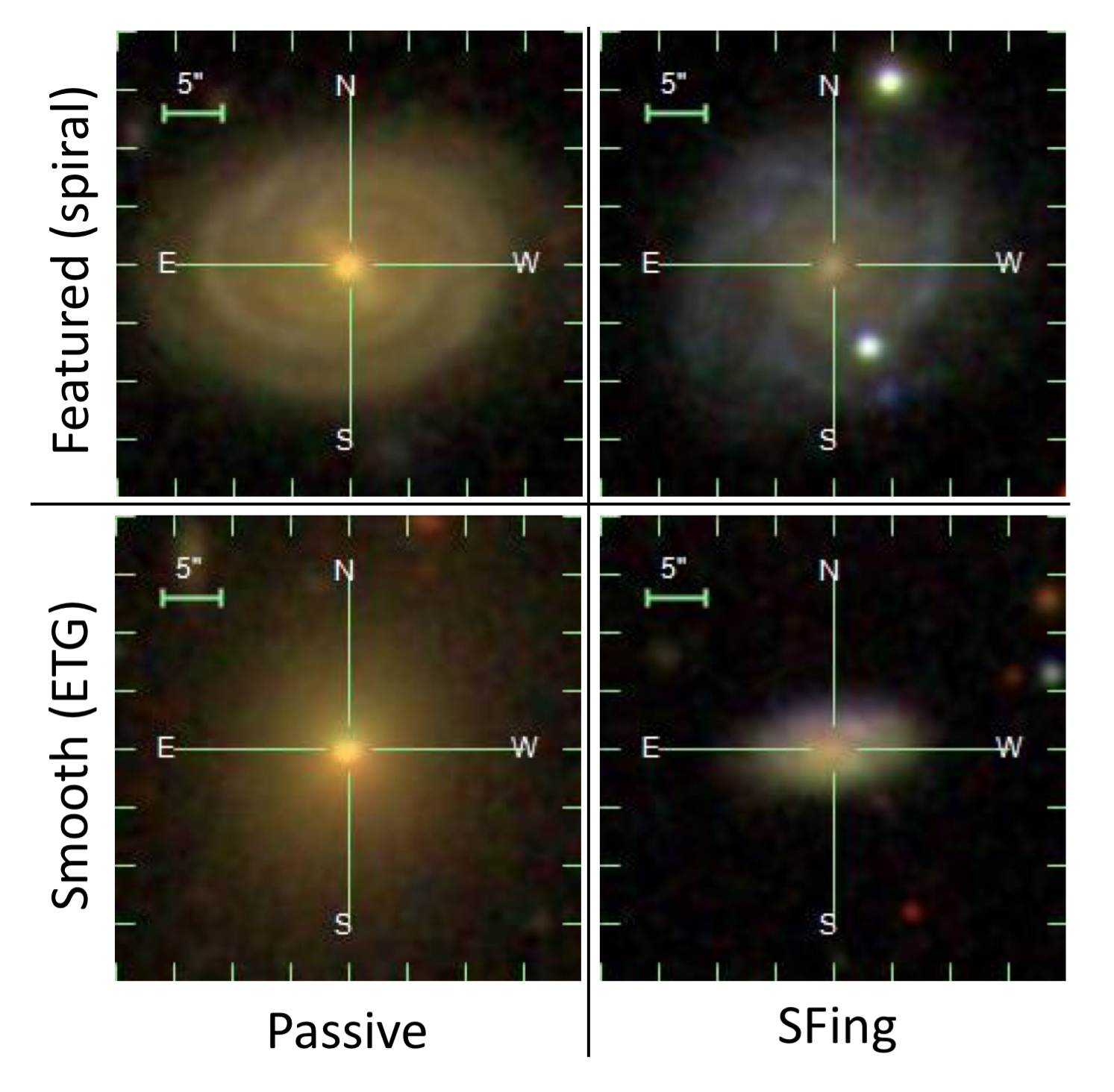}
\caption{\label{fig:examples} Example images ($gri$ colour composites from SDSS, with the scale shown), for two featured (spiral) galaxies (top row), and two smooth galaxies (ETGs; lower row). The left column shows passive galaxies, and the right column shows star-forming galaxies.}
\end{figure}

Morphology information is taken from the Galaxy Zoo 2 Data Release \citep{Willett2013} based on the aggregated classifications of citizen scientists. Each galaxy typically has 44 independent classifications (median count), and these classifications are reduced and aggregated via a careful process that includes weighting to reject low-quality classifications and ``debiasing'' to correct for variations in feature resolution with redshift. A detailed account of these procedures is given in \citet{Willett2013}. The sole morphological quantity we use is hereafter referred to as `$p_{\rm feat}$', which is the consistency-weighted (and redshift debiased) fraction of votes that the galaxy is observed to be featured rather than smooth. This is the first and simplest classification made by Galaxy Zoo volunteers. This parameter can be treated approximately as likelihood that a galaxy is featured -- where $p_{\rm feat}=1$ it is very certain the galaxy has features (most often spiral structure), while $p_{\rm feat}=0$ reveals very featureless smooth early-type galaxies. 
Examples of galaxies with $p_{\rm feat}>0.5$ (featured galaxies, which are almost all spiral galaxies), or $p_{\rm feat}<0.5$ (smooth, or early-type galaxies, ETGs) are shown in Figure \ref{fig:examples}). We also check that no galaxies in our sample contain visual artifacts which may interfere with visual morphological classifications.

Photometric data for this sample is taken from the New York University Value Added Galaxy Catalog (NYU-VAGC) which is a reprocessing of the SDSS imaging optimized for nearby galaxies \citep{blanton_new_2005}. From this catalogue, we use the $(g - r)$ colour and $M_r$ absolute magnitude, which were K-corrected to 0.1 and measured within two times the Petrosian radius. We use the redshift $z$ direct the SDSS Main Galaxy Spectroscopic Survey \citep{abazajian_seventh_2009}. 

Derived quantities such as the stellar mass and sSFR are from the MPA-JHU DR7 release \citep{brinchmann_physical_2004}. Stellar masses provided in the MPA-JHU DR7 release were estimated with photometric fits, with a similar method as \citet{kauffmann_stellar_2003, gallazzi_ages_2005}. For galaxies which are classified as containing an AGN or a low-ionization nuclear emission-line region (LINER) according to a Baldwin, Phillips, and Terlevich (BPT) diagram, the sSFR values are from the technique of \cite{brinchmann_physical_2004}, which finds an empirical relationship between sSFR and the 4000-\AA{} break (D4000). For star-forming galaxies, the sSFR values are determined from the technique in \cite{charlot_nebular_2001}, which uses other emission lines including $H\beta$, [OII], [OIII], [NII], and [SII] to refine the estimation of the sSFR from $H\alpha$ luminosity.

\begin{figure*}
\includegraphics[width=0.9\textwidth]{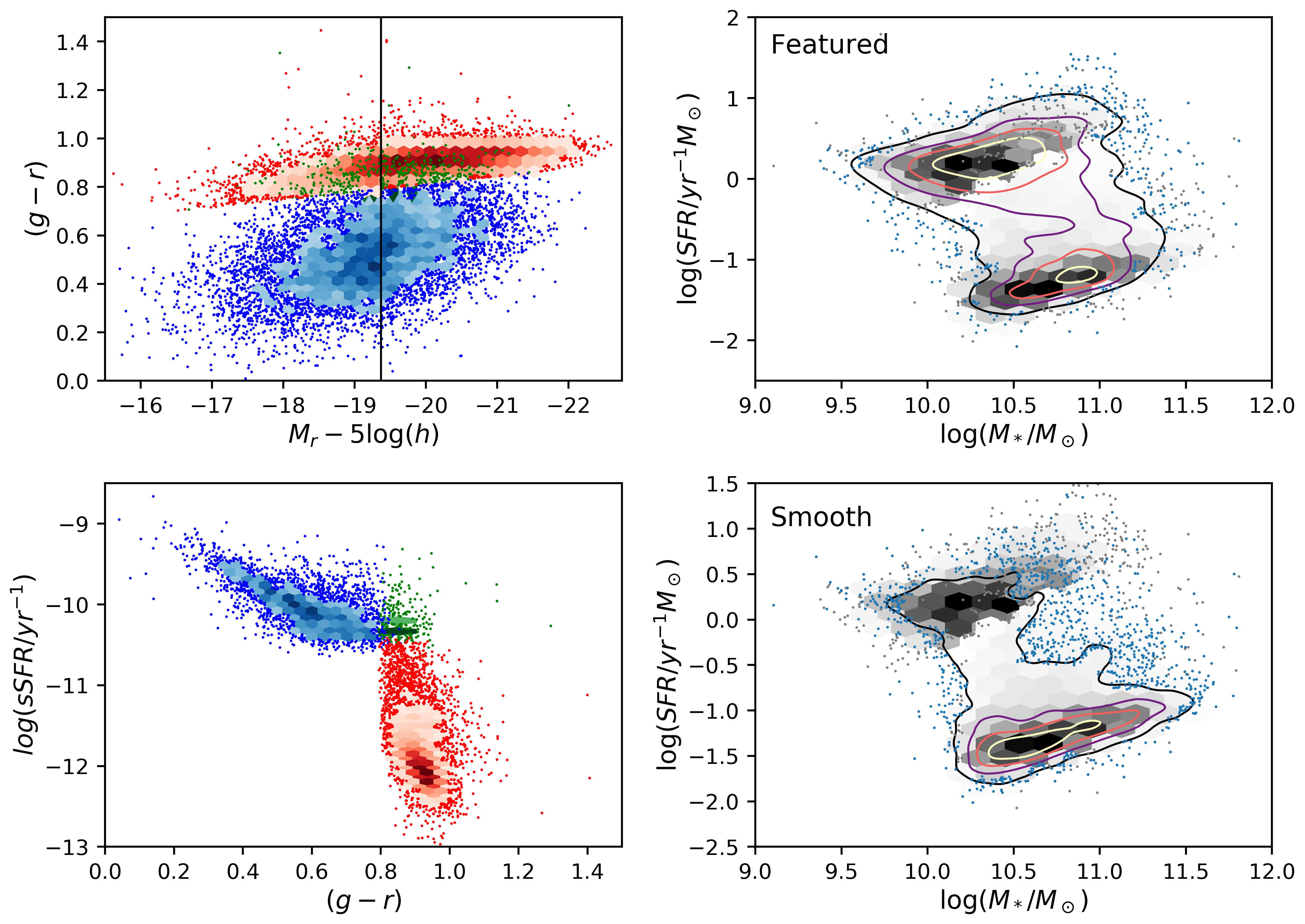}
\caption{\label{sample_plot} Upper left: a colour-magnitude diagram of the parent sample with the magnitude cut of the volume limited sample shown as the vertical line (defined by $M_r -5\log(h) < -19.37$). Lower left: sSFR vs. optical colour for the volume limited sample. In the two left panels, red points and hexagonal bins correspond to red and passive galaxies, blue points and bins correspond to blue and star-forming galaxies, and green points and bins correspond to red and star-forming galaxies. Right panels: the SFR and stellar mass for all galaxies in the volume limited sample. The upper right plot shows featured (spiral) galaxies in contours and blue points, while the lower right plot shows smooth (elliptical) galaxies in contours and blue points. Both plots show the volume limited sample in the grey bins.}
\end{figure*}

In Figure \ref{sample_plot} we provide an overview of the properties of galaxies in our sample. The upper left panel shows the colour-magnitude diagram for the parent sample (i.e. SDSS galaxies with Galaxy Zoo 2 morphologies), and the vertical black line corresponds to our magnitude limit for the volume limited sample which we use in all other plots. The bottom left plot shows the relationship between sSFR and optical $(g-r)$ colour. In both panels, the red points and hexagonal bins show the location of the red, passive galaxies, while blue points and bins correspond to blue and star-forming galaxies; and the green points and bins correspond to optically red, but star-forming galaxies. We use the definitions for colour and star-forming properties previously used by \cite{weinmann_properties_2006}.  The right plots in Figure \ref{sample_plot} show the galaxies in our volume limited sample on a sSFR--stellar mass plot, with the upper plot 
showing featured (spiral) galaxies in the sample, and the lower plot smooth (early-type) galaxies. Our classifications of featured (spiral) and smooth (early-type; ET) are based on $p_{\rm feat} \geq 0.5$ and $p_{\rm feat} < 0.5$ respectively. Our results are similar with more stringent classification requirements of $p_{\rm feat} \geq 0.8$ for featured (spiral) galaxies and $p_{\rm feat} \leq 0.2$ for smooth (ET). In general we observe, as is well known, that featured (spiral) galaxies inhabit the high star-formation region of the plot (sometimes called the star-formation sequence), while the smooth (ET) galaxies occupy the lower star-formation ``passive" region, however there are clearly substantial sub-populations of star-forming early-types and passive spirals. 

To determine group membership, we use the group catalogue from \cite{yang_galaxy_2007}. This catalogue also includes halo mass estimates. These values are calculated with the characteristic luminosity, which is the sum of the group member luminosities adjusted to account for the completeness of the survey. Then, an iterative method estimates the mass to light ratio for a group, and then uses the characteristic luminosity to calculate the halo mass of the group. We exclude any groups which have only two members above our magnitude limit (i.e. pairs) in order to focus on larger groups, leaving a total of 1266 groups, of which 901 (71\%) have smooth (ET) centrals, and 1143 (90\%) have passive centrals. Table \ref{table:cent_type} shows a complete breakdown of the group numbers by the star-formation and morphological properties of the central galaxy, highlighting the now well known result (e.g. \citealt{wolf_stages_2009,Skibba2009,Masters2010,hoyle_fraction_2012}) that passive spirals can make up a significant fraction of both passive galaxies (23\% of passive centrals are spirals in this sample) and spirals (71\% of these massive central spiral galaxies are passive). For example images of these four types of central galaxy see Figure \ref{fig:examples}. Our volume limited sample contains 8230 galaxies in these 1266 groups. 

When plots do not explicitly show halo mass, we use a mass-matching procedure to standardize the mass distributions of different subsamples, following the process of \cite{kawinwanichakij_satellite_2016}. We first create halo mass histograms of the entire sample as well as each subsample (e.g. groups with a star-forming satellite fraction greater than one half). Then we compute a weighting coefficient for each bin:
\begin{equation}
    w_i^{sf(p)} = \frac{N_i}{N_i^{sf(p)}}
\end{equation}
Where $w_i$ is the weighting coefficient for groups in the $i$th bin, $N_i$ is the total number of groups in a bin, and $N_i^{sf(p)}$ is the number of groups with  majority star-forming (or passive) satellites in that bin. Using these weighting coefficients effectively matches the halo mass distributions for subsamples, allowing for comparisons independent of the halo mass.

\begin{table}
\begin{tabular}{l|c|c|c}
                     & Featured (spiral) & Smooth   & All  \\  
                     &  Central & (ETG) Central & \\
                     \hline
SF Central & 105  (8\%)          & 18  (1\%)           & 123 (10\%)   \\ \hline
Passive \\
Central      & 260  (21\%)          & 883 (70\%)          & 1143 (90\%) \\ \hline
All                & 365   (29\%)         & 901  (71\%)         & 1266 
\end{tabular}
\caption{\label{table:cent_type} The number of groups with different types of centrals classified separately by star-formation properties and morphology.}
\end{table}

\section{Results}
\label{results}

\subsection{Galactic Conformity in Star-Formation Properties and Morphology}

In this section, we compare detections of galactic conformity in star-formation properties and in morphology. 

\begin{figure*}
\includegraphics[width=0.9\textwidth]{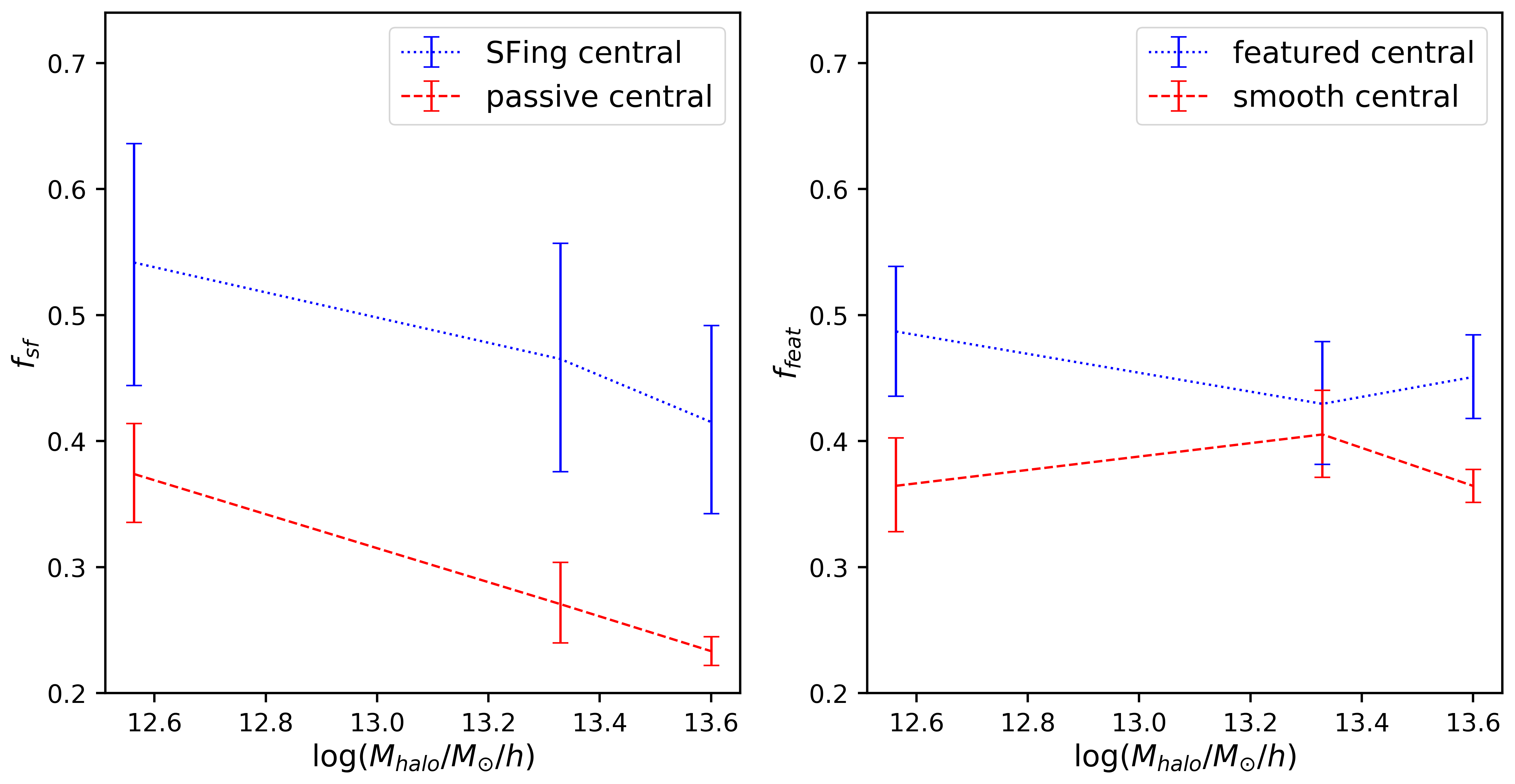}
\caption{\label{fig:mass_frac} Left: the average fraction of star-forming satellites as a function of halo mass. The blue dotted line corresponds to groups with star-forming central galaxies, and the red dashed line corresponds to groups with passive centrals. Right: the same plot but with the fraction of featured satellites. The blue dotted line corresponds to groups with featured centrals, and the red dashed line corresponds to groups with smooth centrals. Errors are calculated with the beta distribution.}
\end{figure*}

We begin by exploring the measure of conformity presented in \citet{weinmann_properties_2006}: in Figure~\ref{fig:mass_frac}, we plot the fraction of satellite galaxies which are star-forming or featured as a function of halo mass. The blue line shows groups with either star-forming or featured (spiral) centrals, while the red line shows groups with either passive or smooth (ET) centrals. The 1 sigma error bars shown are computed with a Bayesian approach using the beta distribution as discussed in \cite{cameron_estimation_2011}. At all halo masses, we detect galactic conformity in star-formation: groups with star-forming or passive centrals have a higher fraction of star-forming or passive satellites. The mean difference in star-forming satellite fraction between groups with star-forming centrals and groups with passive centrals for all halo masses is $f = 0.18\pm$0.08 (i.e. if there are a fraction, $f$ of star-forming satellites observed around a passive central, there will be $f+0.18$ around a star-forming central for the same halo mass). In contrast, we detect galactic conformity in morphology only in the groups with the highest and lowest halo masses, and overall find that the difference in featured satellite fraction for groups with featured (spiral) and smooth (ET) centrals is smaller ($f = 0.08\pm$0.04 averaged over all halo masses). Calculating our errors with jackknife resampling, we find similar, though less significant results with differences of $f = 0.18\pm$0.08 and $f = 0.08\pm$0.06 for star-formation and morphology conformity respectively. {\it Thus we observe a stronger conformity signal for star-formation than morphology.} Another striking difference is that the fraction of star-forming satellites decreases with increasing halo mass, whereas the fraction of featured satellites remains relatively constant with halo mass. These results echo those previously presented in (e.g.) \citet{weinmann_properties_2006} for the star-formation properties, and \citet{hoyle_fraction_2012} for morphology.

\begin{figure*}
\includegraphics[width=0.45\textwidth]{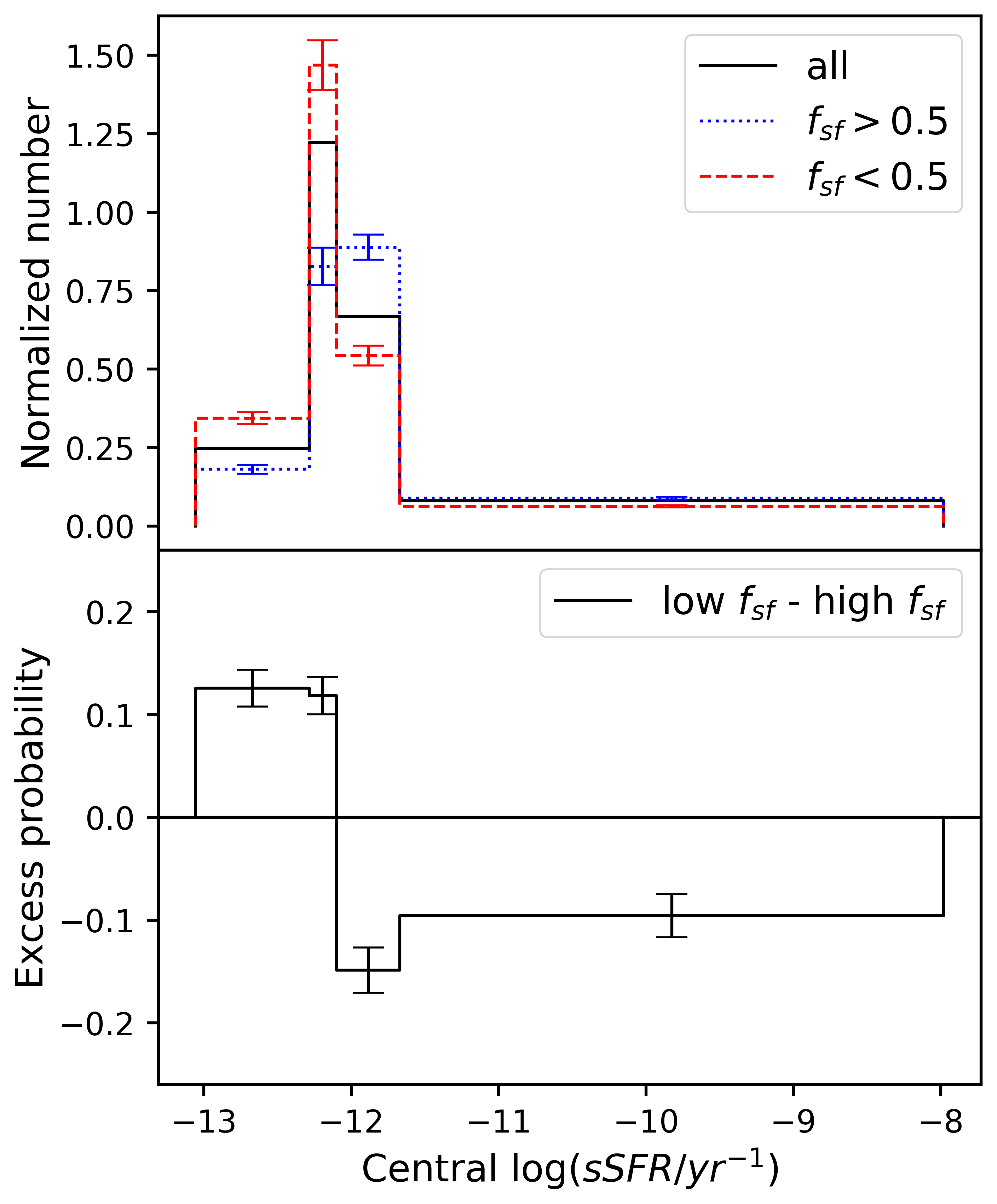}
\includegraphics[width=0.45\textwidth]{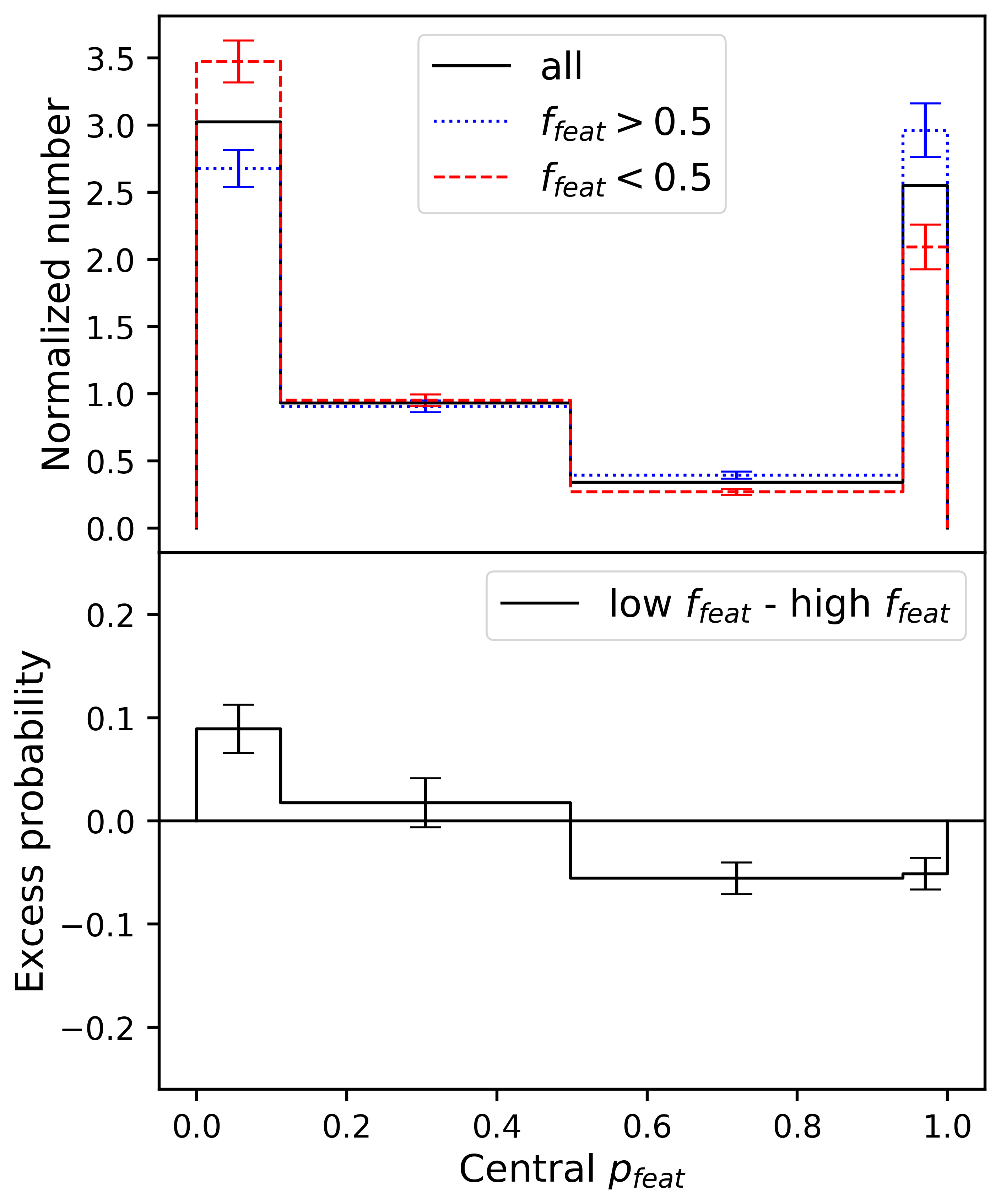}
\caption{\label{conf_sfmorph}Top Left: A normalized histogram of central sSFR in groups. The black solid line corresponds to all groups, the blue dotted line corresponds to groups where more than half the satellites are star-forming, and the red dashed line corresponds to groups where more than half the satellites are passive. Bottom Left: The difference between groups with majority star-forming satellites and groups with majority passive satellites, multiplied by the bin size. Right: The same plots as left, except the $x$ axis is the $p_{\rm feat}$ of the central galaxy of the group, and the red and blue line show groups where over half the satellites are smooth (elliptical) and featured (spiral) respectively.}
\end{figure*}

We now consider a different way to reveal galactic conformity - looking at the normalized number of satellites with different properties relative to the average, as a function of central properties (Figures \ref{conf_sfmorph}, \ref{conf_SF_centmorph} and \ref{conf_morph_centSF}). We define a satellite system as mostly star-forming or featured/spiral (or vice versa) based on the fraction of satellites having that property being $f > 0.5$. When normalized to have an unit area, the differences between these histograms measure the excess probability that a central with a certain sSFR (or $p_{\rm feat}$) will have a satellite system which is mostly star-forming or featured/spiral (or vice versa).

We observe a conformity signal in both star-formation and morphology with these plots (Figure~\ref{conf_sfmorph}, \ref{conf_SF_centmorph}, \ref{conf_morph_centSF}). In all of these figures, the top plots show normalized histograms of all groups (in black solid lines) and the relevant subsamples (in red dashed and blue dotted respectively), while the lower plots show the difference between the subsamples multiplied by the bin width. Thus, the lower row shows the ``excess probability" that a satellite system is in one or other subgroup.  The error bars shown in these plots represent 1 sigma errors (calculated with $\sigma \sim \sqrt{N}$ and normalized appropriately). To remove halo mass as a confounding variable, in Figures~\ref{conf_sfmorph}, \ref{conf_SF_centmorph}, and \ref{conf_morph_centSF} we mass-match each individual histogram to the overall halo mass distribution of groups in the sample, as described in Section~\ref{sample}. 

Starting with Figure~\ref{conf_sfmorph}, the left plots in this figure show the distribution of the central galaxy sSFR. The blue dotted and red dashed lines here indicate the distribution for groups where the the fraction of star-forming satellites is either above or below 50\% respectively.  

We see that groups with lower central galaxy sSFRs have a higher probability of having majority passive satellites, and groups with higher central galaxy sSFRs have a higher probability of having majority star-forming satellites. We also observe that the amount of excess probability increases with more extreme values of central sSFR.

The right side of Figure~\ref{conf_sfmorph} measures morphological conformity -- the central $p_{\rm feat}$ is plotted on the $x-$axis, and the blue and red lines correspond to groups where the fraction of featured (spiral) satellites is greater or less than one half. In these plots, we again see the signal of galactic conformity, now in morphology: groups where the central has high $p_{\rm feat}$ have an excess probability of having mostly featured satellites, and groups where the central has low $p_{\rm feat}$ have an excess probability of having mostly smooth satellites. As was observed in Figure \ref{fig:mass_frac}, morphological conformity has a smaller signal than star-forming conformity.

\subsection{Red Spiral and Blue Elliptical Centrals}

In this section, we examine star-formation and morphological conformity for groups with centrals in a grid of properties (i.e. red and blue spirals and early-types).

\begin{figure*}
\includegraphics[width=0.9\textwidth]{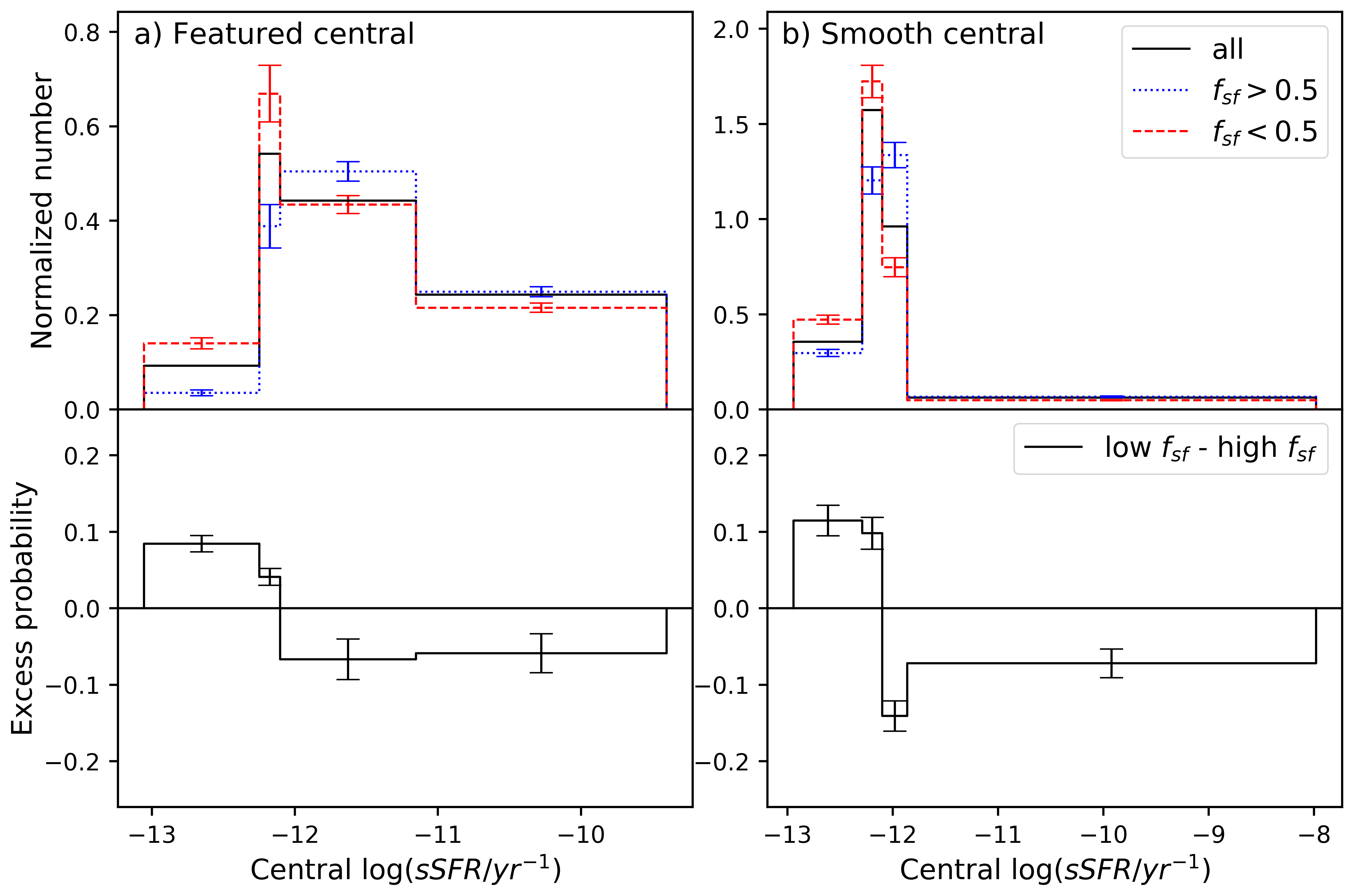}
\caption{\label{conf_SF_centmorph}The left and right columns are split by groups with featured centrals (left) and groups with smooth centrals (right). Top: A normalized histogram of central sSFR in groups. The black solid line corresponds to all groups, the blue dotted line corresponds to groups where more than half the satellites are star-forming, and the red dashed line corresponds to groups where more than half the satellites are passive. Bottom: The difference between groups with majority passive satellites and those with majority star-forming satellites (i.e. the difference between the red and blue lines above), multiplied by the bin size.}
\end{figure*}

Figure~\ref{conf_SF_centmorph} shows star-formation conformity for groups with featured (spiral) centrals (left) and smooth (early-type) centrals (right). 
For both featured and smooth centrals, we see similar levels of star-formation conformity for passive and star-forming centrals. That is the star formation properties of the central do not seem to impact morphological conformity for groups with a spiral central.

\begin{figure*}
\includegraphics[width=0.9\textwidth]{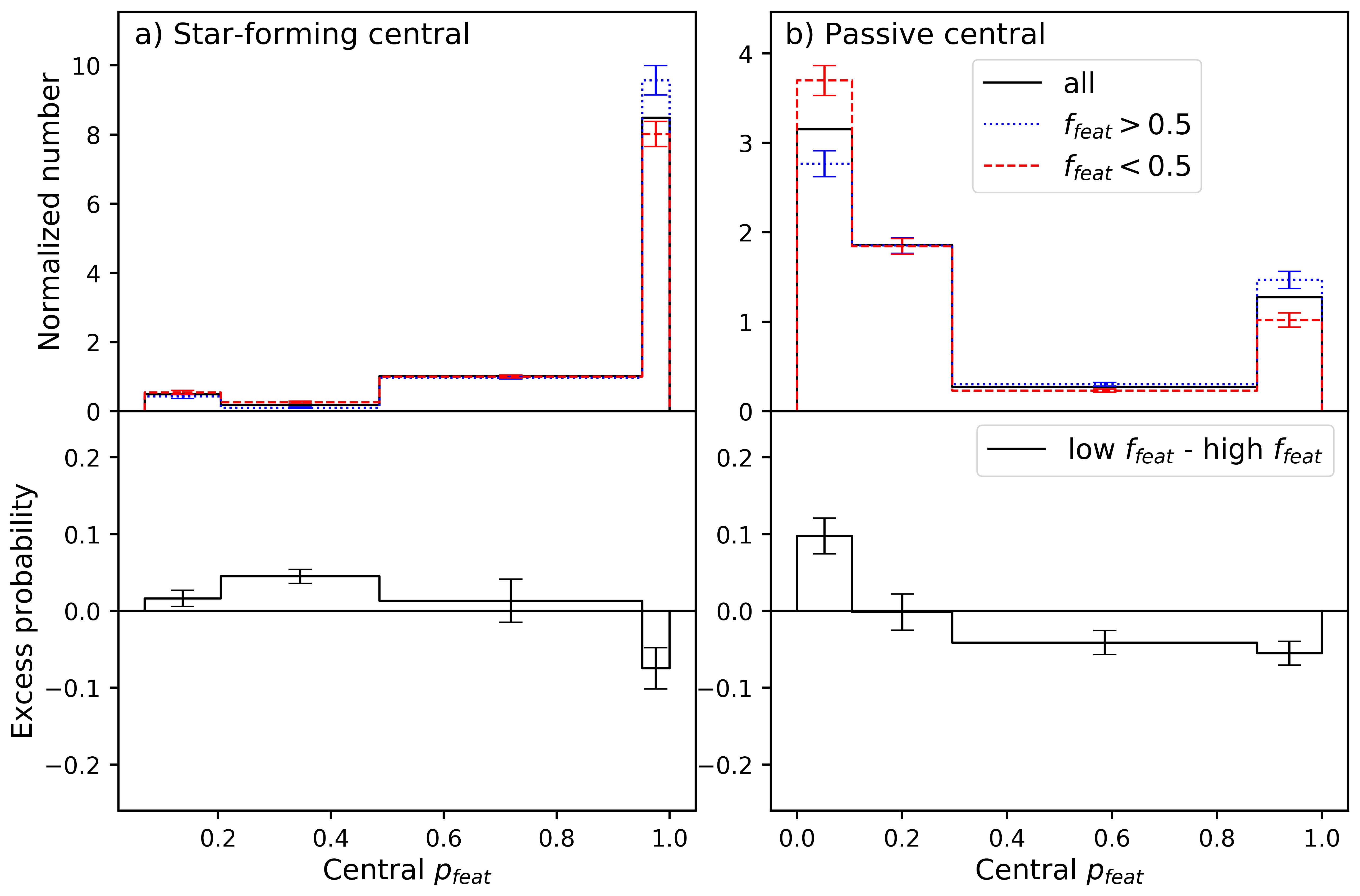}
\caption{\label{conf_morph_centSF}The left and right columns are split by groups with star-forming centrals (right) and groups with passive centrals (left). Top: A normalized histogram of central $p_{\rm feat}$ in groups. The black solid line corresponds to all groups, the blue dotted line corresponds to groups where more than half the satellites are featured, and the red dashed line corresponds to groups where less than half the satellites are featured. Bottom: The difference between groups with less than half featured satellites and those with majority featured satellites (i.e. the difference between the red and blue lines above), multiplied by the bin size.}
\end{figure*}

Figure~\ref{conf_morph_centSF} shows the morphology conformity of groups split by central star-formation properties; the left side shows groups with star-forming centrals and the right side shows groups with passive centrals. We observe greater morphology conformity for groups with passive centrals than those with star-forming centrals. Amongst passive centrals, the conformity signal is of similar strength at low and high $p_{\rm feat}$, thus groups with centrals that are passive spirals and passive ellipticals have similar degrees of morphology conformity. The morphologies of passive central galaxies do not seem to impact star formation conformity for groups with a passive central galaxy.

We detect a morphology conformity signal amongst groups with star-forming spiral centrals, however in the rare groups (they represent $<2$\% of all groups; see Table~\ref{table:cent_type}) with star-forming elliptical centrals, we barely detect a signal. Among star forming central galaxies morphological conformity is only observed for groups with a spiral central galaxy.

\subsection{Discussion of Systematics}\label{sec:systematics}

Galactic conformity has been shown to be sensitive to the details of the group finder. For example \citet{Tinker2018} show that a false conformity signal (a 2-5\% difference in the fraction of quenched satellites, and also the average sSFR of satellites) can be generated if high mass satellites are misclassified as central galaxies, 
which incorrectly creates ``groups''
of entirely satellite galaxies. \citet{campbell_assessing_2015} show that a spurious conformity signal may be observed due to group contamination introduced by group finders, including the group finder of \citet{yang_galaxy_2007} used in this work. They find a weak conformity signal present in mock catalogues containing no conformity due to a randomized shuffle of galaxy properties. 

We use the group catalogue of \citet{yang_galaxy_2007} to identify centrals and satellites. In this catalogue, the central is defined as simply the most massive galaxy in a group, and we limit our sample to groups with N>2 members (so a central and at least two satellites). The catalogue completeness is discussed at length in Section 3.2 of \citet{yang_galaxy_2007}. Figure 2 of that paper shows that 70\% of the Yang et al. groups have interloper fractions less than 10\%; completeness is dependent on halo mass,
and varies between 70-90\% of the groups being 90\% complete. Finally, roughly 70\% of the groups have purity (defined as the number of real members, divided by the number of galaxies identified as members) of 90\% or better. 

As our main results focus on comparing the relative level of conformity between groups with different types of centrals, this effect is not likely to have any major qualitative impact on our results, unless completeness and contamination differs significantly between red/blue and spiral/elliptical centrals. \cite{berti_primus_2019} show some evidence that this may be the case for their probabilistic grouping algorithm, however while they state it may also be present in the \cite{yang_galaxy_2007} groups this has not been investigated.  \cite{yang_galaxy_2007} include the completeness in the region for each galaxy, and we find that the distributions of these completeness values is similar in all of our four subgroups, finding that at a Kolmogorov-Smirnov (KS) test on these distributions cannot reject the hypothesis that they have the same completeness. There is no information on contamination provided in \citet{yang_galaxy_2007} however, since completeness is statistically the same, we expect any impact creating spurious conformity signals should affect all subsets comparably, and so statements about the relative levels of the observed conformity signal should be relatively robust. To test this further would require the mock catalogue technique described in (e.g.) \cite{calderon_small-_2018}, however that scale of analysis is beyond the scope of our short paper.

\section{Summary and Conclusions}
\label{conclusions}

In this paper we present the first investigation of galactic conformity (the tendency of satellites in a group to be more likely share properties with their central galaxy than have different properties) in both star-formation and visual morphological properties separately. Most previous work (e.g. \citealt{weinmann_properties_2006,prescott_galaxy_2011,hartley_galactic_2015}) has conflated these independent observational tracers of a galaxies evolutionary 
history. Our study is therefore uniquely able to identify some interesting signals, which we summarize below. 

The key observational results of this work are:
\begin{enumerate}
    \item \label{one} We detect a galactic conformity signal in both morphology and star-formation properties separately: groups with a star-forming (spiral) central have, on average, a fraction 0.18$\pm$0.08 (0.08$\pm$0.06) more star-forming (spiral) satellites than groups with passive (early-type) centrals at similar halo mass. Thus, morphology conformity is observed to be weaker than star-formation conformity.
    
    \item \label{two} Star-formation conformity is observed to be ubiquitous in groups regardless of the star-formation and morphological properties of the central galaxy.
    
    \item \label{three} Morphology conformity is observed to be strongest in groups with passive centrals (regardless of the morphology of the passive central), is detected in groups with star-forming spiral centrals, and is almost absent in (rare) groups with star-forming early-type centrals.
    
\end{enumerate}

Overall, these results support a physical model where the morphological properties of galaxies are less likely to change (or perhaps change on longer time scales) than star-formation properties in the group environment. This type of behaviour has been previously noted using different observational measures (e.g. \citealt{wolf_stages_2009,Skibba2009}); the new contribution we make in this work is demonstrating that this tendency is also visible in galactic conformity and looking at star-formation and morphological properties independently. Our discovery that star-formation conformity is stronger than morphological conformity (Result \ref{one}) supports models in which the processes changing star-formation properties in groups occur earlier or more frequently than those governing group galaxy morphology. Furthermore, we find evidence (Result \ref{two}) that star-formation conformity is reached independently of the morphology of the central galaxy, while morphological conformity is seen to be dependent on the star-formation properties of the central galaxy (Result \ref{three}). Groups with the highest halo masses tend to have a passive, quenched central galaxy, and therefore begin to assemble earlier (are older) than those with a star-forming central. Thus the increased strength of morphological conformity for groups with passive centrals may indicate that morphological conformity is stronger in older groups. Further study is needed to confirm this, but if true this further supports the idea that the morphological changes necessary to bring about morphological conformity act on longer timescales than the processes which set up star-formation conformity.

Our interpretation is consistent with previous observational work. For example \cite{wolf_stages_2009} find that intermediate and high mass galaxies  ($\log(M_*/M_\odot) \gtrsim 10$) in the in-fall region of clusters undergo morphological change on a slower timescale than the process which quenched their star-formation. Similarly, \cite{Skibba2009} use Galaxy Zoo classifications and determine that satellites are quenched more often than they are morphologically transformed (based on their location on a colour-magnitude diagram).

Looking at simulations,  \cite{tacchella_morphology_2019} recently use IllustrisTNG to study the star-formation and morphological evolution of central galaxies. They quantify morphology by galactic kinematics, specifically the spheroid to total ratio, as well as the stellar mass density profile. They conclude that changes in the star-formation properties of a galaxy are not necessarily accompanied by a change in morphology, and that the morphology of a galaxy is generally determined before the star-formation properties.

Meanwhile, \cite{correa_origin_2019} examine red sequence galaxies in the EAGLES simulation which have undergone a morphological transformation. Their results show that most galaxies move to the red sequence as disc-types. In particular, 55\% of central galaxies in this sample underwent a morphological change after entering the green valley, and while some satellites were found to change morphology first, most change after. 

Under a model where both centrals and satellites form initially as blue spirals, and may eventually transform to red ellipticals, all of these results together appear to paint a consistent picture where morphological properties of galaxies change less frequently (or perhaps more slowly) in the group environment than their star-formation properties. 

Our observations which separate the sample into groups with four types of central galaxy, and the detection of different levels of both star formation and morphological conformity in these would be physically interpreted as follows in this model: 
\begin{itemize}
    \item Most groups (70\% in our sample) have passive ETG centrals, and the majority of the satellites have have both their SF and morphology properties transformed to passive ETGs, hence these groups show both morphological and SF conformity. 
    \item A large fraction of groups (21\%) show centrals where the SF properties of the central have already transformed, but it remains a spiral (red spiral central). In these groups it's also observed that the SF properties, but not the morphologies of most satellites have changed, so these groups shows both morphological and SF conformity. 
    \item In the small fraction of groups where the central galaxy has remained a star-forming spiral (8\%) we observed that most satellites are also star-forming spirals, so these groups also show both morphological and SF conformity.
    \item Very rarely a central galaxy is a star-forming ETG (1\% of groups in our sample). Here we observe that most satellites are also star-forming, but many show spiral structure (so have not morphologically transformed), meaning these groups show only SF conformity. The physical origin of these rare groups is an interesting puzzle. 
\end{itemize}

We note previous works have found that group finding algorithms can lead to spurious conformity detections. \cite{campbell_assessing_2015} use the same \cite{yang_galaxy_2007} group finder as this work and find a weak star-formation conformity signal in a mock catalog created without conformity. \cite{calderon_small-_2018} also find that group finders can introduce a spurious conformity signals in mock catalogs. However, we expect group finding issues of contamination and completeness to be similar among different subgroups (see Section~\ref{sec:systematics}), so comparisons of the magnitude of conformity signals across subgroups are still valid.

While the work presented here makes use of the largest morphological catalogue currently available \citep{Willett2013}, we are stretching the statistics when we separate into groups, and sub-samples on morphology and star-formation properties (particularly for the rarer types of central galaxies, such as star-forming spirals and ETGs). New, larger and deeper surveys which include morphology will allow for more robust statistics and aid in understanding of these complicated correlations.
Future generations of high-resolution wide-area surveys at higher redshift may also provide the sample sizes required to study the properties and evolution of both one-halo \emph{and} two-halo morphological conformity.
Nevertheless our current result clearly reveals that star-formation conformity ought not to be directly interpreted as being due to morphological conformity/changes. Morphology and star-formation properties of galaxies are strongly correlated; however they are independent measures of the physical state of galaxies and should be treated as such, or we lose vital information on how galaxies in our Universe evolve.

\section*{Acknowledgements}
This publication has
been made possible by the participation of more
than 200,000 volunteers in the Galaxy Zoo project. Their contributions are individually acknowledged at http://authors.galaxyzoo.org. Galaxy Zoo 2 was developed with the help of a grant from
The Leverhulme Trust. 
We acknowledge useful comments, particularly on early observations related to galactic confirmity, from W. C. Keel. 
Funding for the SDSS and SDSS-II was provided by the Alfred P. Sloan Foundation, the Participating Institutions, the National Science Foundation, the U.S. Department of Energy, the National Aeronautics and Space Administration, the Japanese Monbukagakusho, the Max Planck Society, and the Higher Education Funding Council for England. The SDSS Web Site is http://www.sdss.org/.
BDS acknowledges support from the National Aeronautics and Space Administration (NASA) through Einstein Postdoctoral Fellowship Award Number PF5-160143 issued by the Chandra X-ray Observatory Center, which is operated by the Smithsonian Astrophysical Observatory for and on behalf of NASA under contract NAS8-03060.

This research made use of \textsc{astropy}, a community-developed core \textsc{python} ({\tt http://www.python.org}) package for Astronomy \citep{Robitaille2013}; \textsc{matplotlib} \citep{Hunter:2007}; \textsc{numpy} \citep{Walt2011}; and \textsc{topcat} \citep{2005ASPC..347...29T}. We also made use of {\tt https://github.com/CKrawczyk/densityplot}.

Galaxy Zoo 2 data are available at {\tt data.galaxyzoo.org}. Spectrum measurements from the MPA-JHU are available at {\tt https://wwwmpa.mpa-garching.mpg.de/SDSS/DR7/}. SDSS data are available at {\tt sdss.org}

\bibliographystyle{mnras}
\bibliography{galactic_conformity}

\bsp
\label{lastpage}
\end{document}